\documentclass[aps,10pt,twocolumn,superscriptaddress]{revtex4-1}
\usepackage[colorlinks=true,allcolors=blue]{hyperref}
\usepackage{amssymb}
\usepackage{amsmath}
\usepackage{graphicx}
\usepackage[caption=false]{subfig}
\usepackage{amsfonts}
\usepackage{xcolor}
\usepackage{epsfig}
\usepackage{color}
\usepackage{bm}
\usepackage{tabularx}
\usepackage{multirow}
\usepackage{mathtools}
\usepackage{xfrac}
\usepackage{blkarray}
\usepackage{bbold}
\usepackage[mathscr]{euscript}
\usepackage{autobreak}
\usepackage{comment}
\usepackage[utf8]{inputenc}
\usepackage{newunicodechar} 
\newunicodechar{̈}{\\"} 

\newcommand{\bracket}[2]{\langle{#1}\vert{#2}\rangle} 

\newcommand{\vb}{V$_{\rm B}$}

\usepackage{soul}
\graphicspath{{./Figures/}}

\begin{document}
\title{Optimizing the depth-dependent nitrogen-vacancy center quantum sensor in diamane}

\author{Pei Li}%
\thanks{These authors contributed equally to this work}
\affiliation{Beijing Computational Science Research Center, Beijing 100193, China
}%
\affiliation{School of Integrated Circuit Science and Engineering, Tianjin University of Technology, Tianjin 300384, China}

\author{Guanjian Hu}%
\thanks{These authors contributed equally to this work}
\affiliation{Beijing Computational Science Research Center, Beijing 100193, China
}%

\author{Xiao Yu}%
\affiliation{Institute of Electronic Engineering, China   Academy of Engineering Physics, Mianyang, Sichuan 621000, China
}%

\author{Bing Huang}
\affiliation{Beijing Computational Science Research Center, Beijing 100193, China
}%
\affiliation{School of Physics and Astronomy, Beijing Normal University, Beijing 100875, China}

\author{Song Li}%
\email{li.song@csrc.ac.cn}
\affiliation{Beijing Computational Science Research Center, Beijing 100193, China
}%

\date{\today}
\begin{abstract}
Negatively charged nitrogen-vacancy (NV) center in diamond is the representative solid state defect qubit for quantum information science, offering long coherence time at room temperature. To achieve high sensitivity and spatial resolution, shallow NV center near the surface are preferred. However, shallow NV center suffers from surface states and spin noise which reduce the photostability and coherence time. In this work, we systematically study the NV center in recently reported two-dimensional diamond, known as diamane---using first-principles calculations. We show that the quantum confinement in finite-layer diamane, with appropriate surface passivation, could significantly modify the band structure. In particular, we identify oxygen surface termination capable of hosting NV center in diamane while optimizing photostability compared to bulk diamond. Furthermore, layer-dependent NV center demonstrates tunable zero-phonon-line and suppressed phonon side band, while retaining long coherence time. Our findings highlight diamane as a promising platform for NV-based quantum information processing with improved optical properties and stability.     
\end{abstract}

\maketitle

%
%
\section{Introduction}
The NV center in diamond is a prominent solid state qubit for quantum information processing which could operate at ambient environment. It has been employed for sensing magnetic~\cite{maze2008nanoscale, casola2018probing}, electric fields~\cite{dolde2011electric,balasubramanian2008nanoscale} and exotic physical phenomenons at extreme conditions in nanoscale~\cite{lesik2019magnetic,schirhagl2014nitrogen}. The quantum operation protocols of the NV center rely on its negative charge state (NV$^-$), which can be achieved by capturing one extra electron from local environment, e. g. nitrogen substitution. NV$^-$ has a triplet ($S = 1$) ground state and the zero field splitting (ZFS) is 2.87 GHz. The ZFS offers the spin coherent control between $m_s$ = 0 and $m_s$ = $\pm1$ sublevels by resonant microwave. There exists a spin selective decay path for $m_s$ = $\pm1$ sublevel, enabling optical initialization and readout of the spin states through spin-dependent fluorescence. 

To achieve higher sensitivity and nanoscale spatial resolution, the NV center should be placed near the diamond surface with proper functionalized terminations~\cite{janitz2022diamond,nagl2015improving}. In such cases, however, the diamond surface can compromise charge stability~\cite{hauf2011chemical}, optical properties~\cite{neethirajan2023controlled}, and spin coherence of NV center ~\cite{kim2015decoherence} due to factors such as non-optimal crystal orientation~\cite{michl2014perfect}, surface spin noise~\cite{rosskopf2014investigation}, and unwanted surface states intrusion~\cite{kaviani2014proper}. 
Several studies investigated the passivation for various dimaond facets---including (001), (011), (111), and (113)---to saturate dangling bonds and minimize the surface states which can cause NV$^-$ blinking or bleaching~\cite{kaviani2014proper,chou2017nitrogen,li2019oxygenated}. 
While first-principles calculations have identified ideal surface terminations for NV$^-$ operation, experimentally achieving controlled functionalization of bulk diamond surfaces---via plasma, acid treatment, annealing, and related methods---remains challenging. For example, oxidation of diamond surface could stabilize NV$^-$ and diminish surface states, yet excessive oxidation may induce additional paramagnetic surface defects, degrading spin coherence~\cite{sangtawesin2019origins}. Another intrinsic limitation of NV$^-$ center is that its zero-phonon-line (ZPL) only contributes 3\% to the total emission, result in a broad phonon side band (PSB). This makes the ZPL is difficult to observe at room temperature, and the optical signal of single NV center is relatively dim. These have motivated the design and integration of optical cavities on diamond to enhance the optical emission of NV$^-$ center~\cite{jensen2014cavity}. However, the exceptional hardness of diamond poses significant challenges for micro- and nanofabrication of optical cavities. The current situation motivates further optimization of NV$^-$ center.

Compared with bulk materials, two-dimensional (2D) wide-bandgap semiconductors offer versatile functionalities and support tunable growth techniques. In recent years, single photon emitters (SPEs) have been extensively reported in various 2D semiconductors, such as hexagonal boron nitride~\cite{sajid2020single} and transition metal dichalcogenides~\cite{koperski2015single}. Although the exact atomic structures of some emitters remain unclear, SPEs in 2D semiconductors feature high quantum efficiency and brightness. These advances raise the question of whether NV$^-$ in diamond can be realized in a 2D form. Recently synthesized ultrathin diamond films, known as diamane~\cite{sorokin2021two,bakharev2020chemically}, exhibit unique physical properties and chemical stability with proper surface passivation. This platform could potentially overcome the inherently low quantum efficiency of NV$^-$ in bulk and provide additional functionalities arising from reduced lattice stiffness. Motivated by this, in this paper, we systematically investigate the magneto-optical properties of NV center in diamane using first-principles calculations. The paper is organized as follows: Sec.~\ref{sec1} presents the electronic properties of few-layer diamane, including the bandgap, electron affinity, and surface states. Sec.~\ref{sec2} discusses the layer-dependent magneto-optical characteristics of NV$^-$. Finally, Sec.~\ref{sec:discussion} provides a broader discussion of the implications of our findings.

\begin{figure*}[tb]
\includegraphics[width=2\columnwidth]{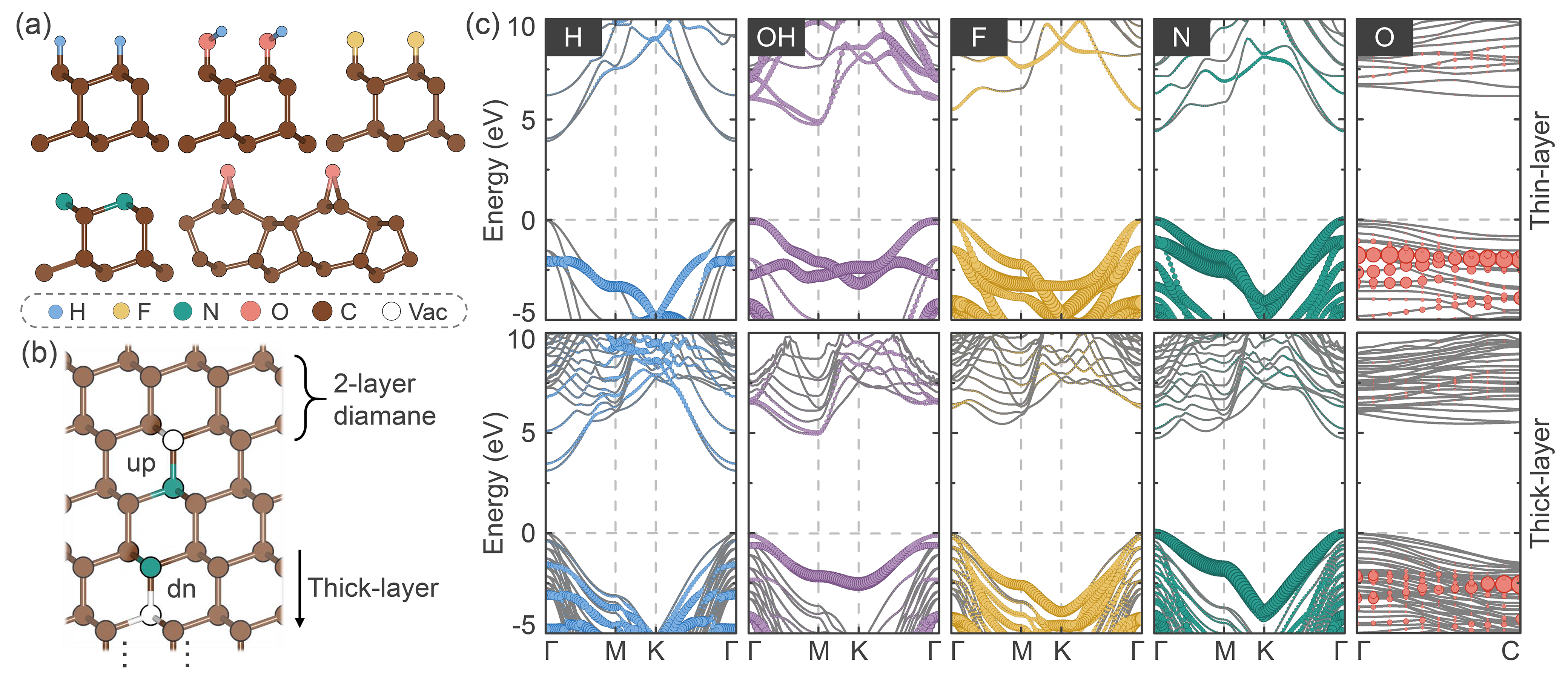}
\caption{\label{Figure1}%
\textbf{a} The surface terminations considered in this paper. We use the 2 $\times$ 1 reconstructed model for O-termination. \textbf{b} Schematic side view of the diamane slab along (111) direction. The two configurations of NV$^-$ are shown. \textbf{c} The HSE band structures of diamane slab with surface terminations. The color dots indicate the projected surface states from terminations. The upper figures are band structures of thin-layer and the bottom figures are band structures of thick-layer.}
\end{figure*}

\section{Results}
\subsection{\label{sec1}Electronic structures of passivated diamane}
Few-layer diamane without surface passivation is theoretically unstable and tends to undergo graphitization, decomposing into multilayer graphene~\cite{kvashnin2014phase}. The reverse process---diamondization of multilayer graphene---can be achieved through external pressure or chemical surface passivation. Notably, hydrogenated and fluorinated diamane have been experimentally synthesized under ambient pressure conditions and characterized via visible and ultraviolet Raman spectroscopy~\cite{piazza2019low,piazza2020raman,son2020tailoring}. Building on these advances, it is feasible to incorporate nitrogen atoms during the fabrication process to form NV center in diamane at controlled concentrations.
Hence, we first perform a comprehensive investigation of the electronic structure of surface-passivated diamane with varying thickness, as illustrated in Fig.~\ref{Figure1}. The model considered is (111)-oriented, with the number of bilayers ranges from 2 to 9. Five types of surface terminations---H, F, O, OH, and N---are examined, as these species commonly occur on bulk diamond surfaces. The NV$^-$ center is placed perpendicularly to the surface, with its symmetry axis aligned along with the normal vector of the (111) plane. This axial alignment not only enhances the optical collection efficiency~\cite{neu2014photonic}, but also preserves the $C_{3v}$ symmetry of the NV$^-$ center, thereby minimizing the splitting of the degenerate $e$ orbitals. Due to the reduced dimensionality of the system, surface terminations have a pronouced influence on the electronic structure, as evidenced by the projected density of states (PDOS) of the terminating atoms shown in Fig.~\ref{Figure1}c.   

\begin{figure*}[tb]
\includegraphics[width=2\columnwidth]{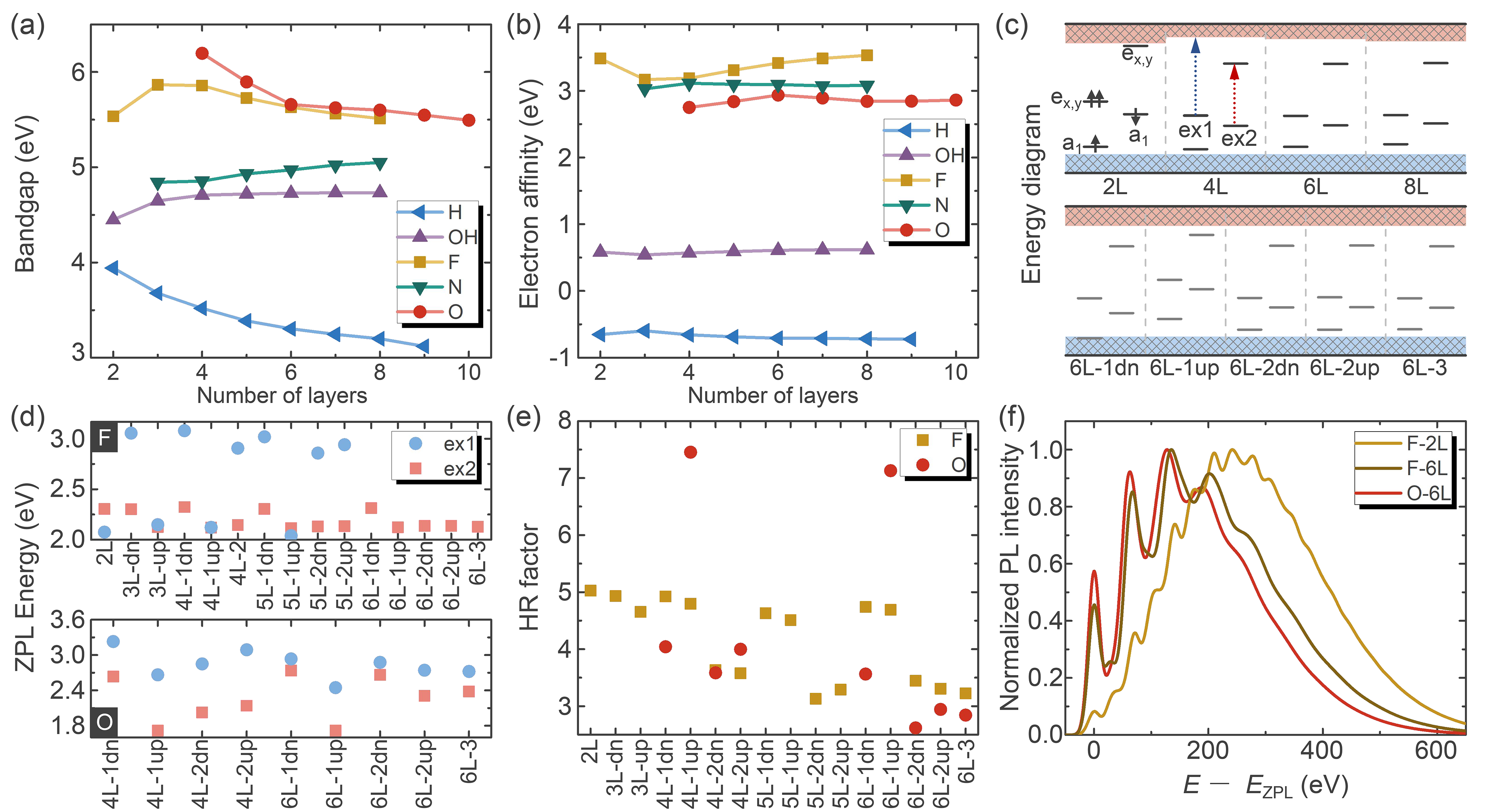}
\caption{\label{Figure2}%
\textbf{a} The bandgap evolution as increase of number of layers with surface terminations considered in this paper. \textbf{b} The electron affinity evolution as increase of number of layers with surface terminations. \textbf{c} The schematic diagram of energy levels of NV$^-$ in F-diamane with different thickness. Here the position of NV$^-$ is in the middle of slab. \textbf{d} The calculated ZPL energies of F-diamane and O-diamane. The ex1 and ex2 optical transitions are considered. The $x$ axis indicates the position and configuration of NV$^-$. (e) The HR factor of F-diamane and O-diamane. (f) The simulated PSB aligned to ZPL position. }
\end{figure*}

Hydrogen termination (H-diamane) significantly modifies the bandgap relative to bulk diamond. Previous investigations have defined the states just below the conduction band minimum (CBM) as ``surface states", originating from the surface terminations and the topmost layer carbon atoms~\cite{kaviani2014proper}. In ultrathin diamane, this conventional definition of surface state is unsuitable since the system itself is surface per se. The so-called surface states can strongly hybridize with the empty defect levels of the NV$^-$ center. This interaction may facilitate excitation-induced electron transfer to the vacuum due to the presence of a negative electron affinity (NEA), leading to a charge-state conversion from NV$^-$ to NV$^0$, thereby quenching its optical signal --- an effect observed experimentally~\cite{bradac2010observation}. 
Our calculations show that these surface-derived states are robust, persisting regardless of the H-diamane thickness or the depth of the NV$^-$ center. As shown in Fig.\ref{Figure2}a, the bandgap of H-diamane decreases with increasing layer count. Although the thinnest 2-layer H-diamane retains a bandgap of 3.94~eV, the conduction band still energetically overlaps with the unoccupied $e$ states of the NV$^-$. In all cases, the NEA is preserved, although its magnitude is smaller than the experiment value due to the finite-size effects~\cite{cui1998electron}. Notably, the surface states in H-diamane are delocalized above the surface with strong dispersion, known as Rydberg states~\cite{kaviani2014proper}. Together with the NEA, these features make the H-diamane act as an electron emitter, which is detrimental for NV$^-$-based sensing applications.

Further oxidation of the H-diamane can produce hydroxyl (OH) groups on diamond, as observed in experimentally~\cite{pehrsson2000oxidation,yoshida2018formation}. The calculated bandgap of 2-layer OH-diamane is 4.45~eV, gradually converging to 4.73~eV as the layer count increases. The surface exhibits a PEA of approximately 0.6~eV. Although the surface states near CBM are shallower than those in H-diamane, they still have the possibility to mix with empty states of NV$^-$ center, potentially affecting charge stability. 

In addition to H- and OH-diamane, fluorine termination (F-diamane) is another frequently employed surface configuration. The bandgap of F-diamane shows a non-monotonic dependence on layer count. In the 2-layer structure, F atoms contribute significantly to CBM, whereas the bandgap is close to the bulk value (5.5~eV). With increasing layers, the contribution of F atoms to the CBM gradually diminishes, accompanied by a widening of the bandgap. With increased thickness, the electronic structure begins to resemble that of bulk diamond, and the bandgap subsequently shows a decreasing trend. 
The relatively large bandgap provides well-isolated defect states of NV$^-$ center, effectively preventing hybridization with surface states. The calculated PEA is above 3~eV, which may suppress direct photoionization of NV$^-$. However, in ultrathin cases such as 2-layer F-diamane, residual surface states may trap electrons via two-phonon absorption under high-power laser excitation~\cite{siyushev2013optically}, potentially compromising the stability of the optical signal. Furthermore, the non-zero nuclear spin of fluorine atoms ($I = 1/2$) could introduce local magnetic noise, contributing to decoherence of NV$^-$ center near the surface. 



Another surface configuration is nitrogen termination (N-diamane), which has been experimentally realized on diamond surfaces~\cite{stacey2015nitrogen,attrash2021nitrogen}. 
The nitrogen-terminated surface forms atomically smooth C–N–C bridge structures, distinct from the previously discussed terminations. This configuration yields a bandgap of approximately 5~eV and a PEA exceeding 3~eV.
The observed bandgap reduction compared to bulk diamond is attributed to finite-size effects, and the gap gradually increases with the number of layers. Although N-diamane provides a clean bandgap and favorable PEA~\cite{chou2017nitrogen}, the non-zero nuclear spin of N atoms can also introduce magnetic noise near the surface, potentially limiting NV$^-$ spin coherence. 

Finally, we examine oxygen termination (O-diamane), a widely adopted approach for surface functionalization of diamond. Due to the presence of a single dangling bond on each surface carbon atom, oxygen exhibits high surface reactivity, often leading to surface reconstruction. 
Here we consider the (2 $\times$ 1) reconstructed surface with epoxy-like oxygen arrangement, as observed experimentally~\cite{loh2002oxygen}. Due to the increased computing time, we just plot the HSE band structure along the path containing the VBM and CBM. The projected density of states reveals that oxygen-derived states reside deep within the valence and conduction bands. As the number of layers increases, the bandgap decreases from 6.20 to 5.49~eV. Combined with its favorable PEA, these features make O-diamane a promising host for NV$^-$ center. In the following section, we investigate the magneto-optical properties of NV$^-$ center embedded in diamane. 

\subsection{\label{sec2}Optical properties of NV$^-$ center in diamane}
We model the NV$^-$ center in large supercells with various layer numbers and two distinct configurations---``up" and ``down"---defined by the relative position of the nitrogen atom and the adjacent vacancy, as illustrated in Fig.~\ref{Figure1}b.
Taking F-diamane as an example, we investigate both the size and surface effects on the defect levels and bandgap. As shown in Fig.~\ref{Figure2}c, these effects affect the defect states and bandgap in ultrathin structures,  but diminish rapidly with increasing layer number. 
In bulk diamond, the lowest optical excitation of the NV$^-$ center occurs between the $a_1$ and degenerate $e$ states in the spin-down channel (denoted as ex2), corresponding to a ZPL energy of 1.95~eV. In contrast, in 2-layer F-diamane, the surface-induced shifts of defect levels reduce the energy separation between the $e$ states and the CBM in the spin-up channel (denoted as ex1), potentially making ex1 the lowest-energy optical transition.
We calculate the excitation energies of these two transitions, as shown in Fig. ~\ref{Figure2}d. 
We find that, across all thicknesses, ex1 consistently exhibits a lower ZPL when NV$^-$ is located at the topmost layer. Specifically, in 2-layer F-diamane, the ZPL for ex1 is 1.89~eV, which is lower than the 2.16 eV observed for ex2. This suggests that under certain conditions, the optical excitation may follow the ex1 pathway rather than the desired bound-bound ex2 transition. 
Although this transition does not fully neutralize NV$^-$ center, the excited electron may be driven away temporarily, leading to blinking phenomena---especially under high-power laser excitation via two-photon absorption. This effect is strongly dependent on excitation intensity. Our calculations indicate that such an altered excitation pathway occurs only when the NV$^-$ center is located immediately below the surface termination. Placing the NV$^-$ center deeper within the structure effectively suppresses it. 
We further simulate the PSB of the conventional ex2 transition. When the NV$^-$ center is placed in the top layer, the PSB broadens substantially. The calculated Huang–Rhys (HR) factor increases to approximately 5, compared to 3.67 in bulk diamond, resulting in a Debye–Waller (DW) factor of merely ~0.7\%. This significant reduction in DW factor implies a further drop in quantum efficiency.
Overall, these results show that the 2-layer F-diamane is not a suitable host for NV$^-$. Moreover, the NV$^-$ should not be placed in the top layer of diamane if the desired optical excitation is to be preserved. 

Oxygenation provides a favorable surface to host shallow NV$^-$ center in diamane. Although the NV$^-$ centers located at the topmost layer of O-diamane undergo noticeable structural distortion, the desired excitation pathway (ex2) remains energetically lower than the undesired spin-up excitation (ex1), as shown in Fig.~\ref{Figure2}d. The ZPL energy of surface-proximal NV$^-$ centers can vary substantially, ranging from 1.32 to 2.23~eV depending on their specific configuration. However, these centers also exhibit pronounced phonon coupling: the calculated HR factor reaches up to 7.45 (see Fig.~\ref{Figure2}e), far exceeding the bulk value (3.67), resulting in reduced quantum efficiency. 
Furthermore, in these configurations, the symmetry axis of the NV$^-$ center is misaligned with respect to the surface normal, which can lower photon collection efficiency in practical applications. In general, both the ZPL energy and HR factor of NV$^-$ centers at the top surface deviate from bulk values but gradually converge as the NV$^-$ center is embedded deeper into the slab, as shown in Figs.~\ref{Figure2}d and \ref{Figure2}e. Interestingly, in 6-layer O-diamane, the HR factor for an NV$^-$ center located at the mid-plane of the slab decreases to approximately 2.8, which is even lower than the bulk reference. This reduction may be attributed to constrained lattice relaxation and limited atomic displacement in few-layer systems due to reduced dimensionality. Such behavior suggests a promising strategy for optimizing the optical performance of the NV$^-$ center via controlled tuning of diamane thickness and surface passivation chemistry. 

\begin{figure}[tb]
\includegraphics[width=\columnwidth]{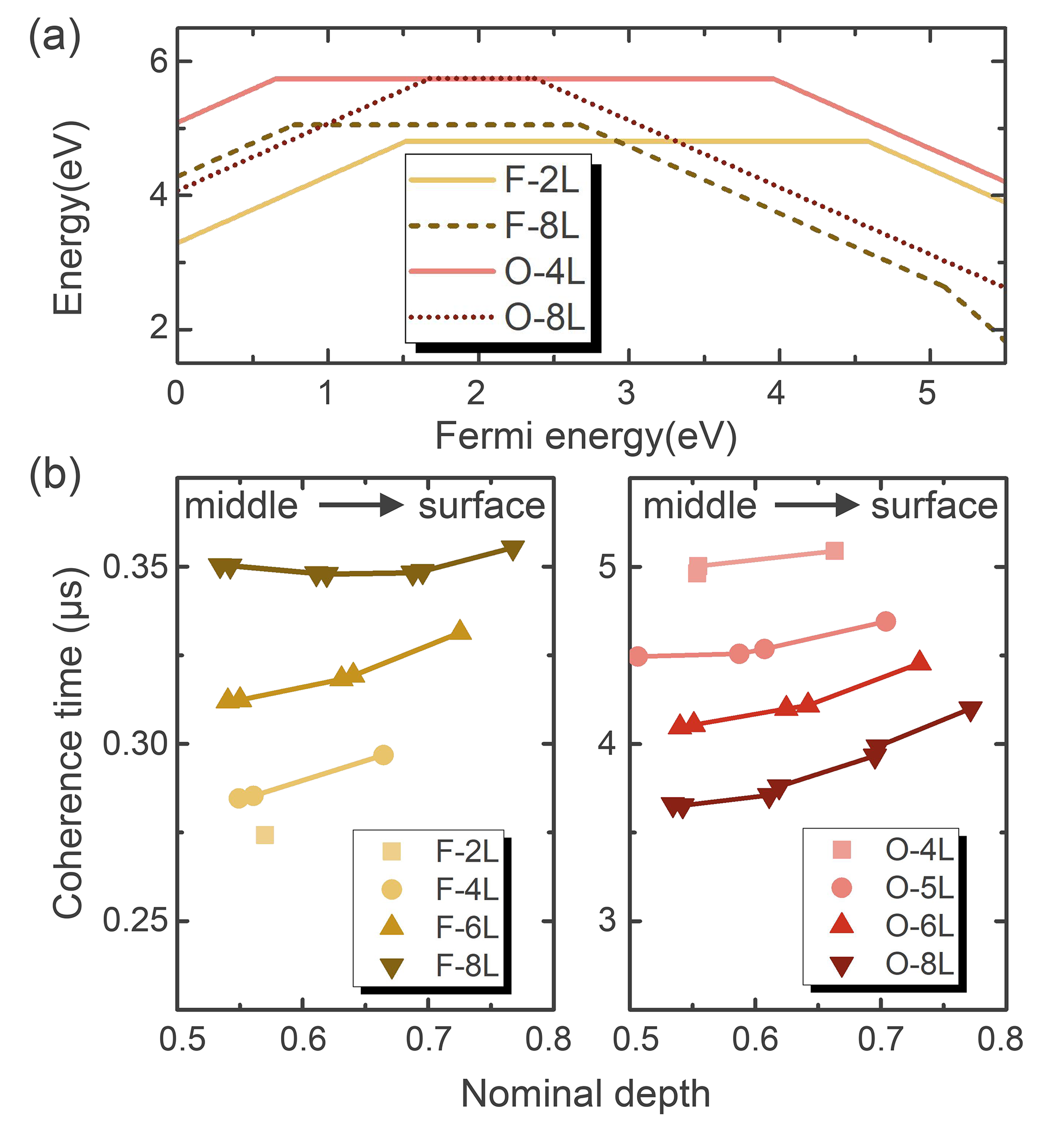}
\caption{\label{Figure3}%
\textbf{a} The formation energy of F- and O-diamane. \textbf{b} The calculated coherence time of NV$^-$ as a function of depth.}
\end{figure}

We further calculate the formation energies and charge transition levels (CTLs) of NV centers in both F- and O-terminated diamane in Fig.~\ref{Figure3}a. Compared to bulk diamond, the formation energies of NV$^0$ centers in few-layer diamane are significantly reduced, and the corresponding CTLs exhibit pronounced shifts due to surface and confinement effects. 
In 2-layer F-diamane, the NV$^0$ is thermodynamically favored over a wide range of Fermi level within the bandgap. The CTL $\varepsilon$(0/$-$1) is at 4.59 eV, substantially higher than the bulk reference value of approximately 2.78~eV. This implies that the NV$^-$ is only stabilized under heavy $n$-type doping conditions, and is otherwise prone to ionization into NV$^0$. Increasing thickness, the CTLs gradually shift toward bulk-like values. For example, in 8-layer F-diamane, NV$^-$ becomes stable within a Fermi level window of 2.67–5.09~eV, indicating restored charge-state stability. Compared to F-diamane, O-terminated diamane exhibits improved stabilization of the NV$^-$ charge state. In 2-layer O-diamane, the $\varepsilon$(0/$-$1) transition level is calculated to be 2.37~eV, which is already close to the bulk value. From the perspective of charge-state stability, these results suggest that the NV center should be placed at least 8~\AA\ beneath the surface to suppress ionization and maintain the NV$^-$ state under experimentally relevant conditions. 

Another important parameter of NV$^-$ is the ZFS. The first-order ZFS due to spin-spin interaction contains a longitudinal component of the magnetic dipole-dipole interaction ($D$) and a transverse part ($E$). The Hamiltonian can be expressed as:
\begin{equation}
H_{ZFS} = D[S^2_z-1/3S(S+1)]+E(S^2_x-S^2_y)\text{,}
\end{equation}
the $D$ and $E$ could be calculated through the diagonalized $\hat{\textbf{D}}$ tensor with $D$ = 3/2$D_{zz}$ and $E$ = ($D_{yy}$ - $D_{xx}$)/2. In bulk diamond under ideal conditions (i.e., in the absence of external fields or lattice distortion), the transverse component $E$ is nominally zero due to the $C_{3v}$ symmetry of the NV center. Using the PBE functional and without applying spin-decontamination, we obtain a ZFS value of $D = 2.99$~GHz for NV$^-$ in bulk diamond, in agreement with previous theoretical results~\cite{gali2019ab, loubser1978electron}. For diamane, we adopt high-symmetry NV configurations to minimize symmetry-breaking artifacts. As a result, the calculated $E$ remains below 3~MHz when the NV center is placed sufficiently far from the surface, suggesting minimal transverse distortion.
In 2-layer F-diamane, $D$ = 2.95 GHz, which is comparable to the bulk value, while $D$ increases to 3.07 GHz in 6-layer F-diamane. 
Notably, in ``up” configurations, where the nitrogen atom is closer to the surface, the increase in $D$ is more pronounced than in ``down” configurations. This is attributed to the fact that the the major contribution to $D$ arise from the three carbon atoms neighboring the vacancy site, which are more affected by the surface environment in ``up” cases.
For OH-terminated diamane, the ZFS decreases to $D = 2.86$~GHz in the 2-layer structure, highlighting the sensitivity of NV wavefunctions and local geometry to surface chemical environments.
The most significant deviation is observed in 2-layer O-diamane, where surface reconstruction induces substantial distortion of the NV site. Here, $D$ decreases to 2.1~GHz and the transverse component $E$ becomes considerable (up to 394 MHz). However, both $D$ and $E$ recover to near-bulk values as the NV$^-$ center is placed deeper within the slab.
These findings demonstrate that the ZFS of NV$^-$ centers in diamane can be modulated via surface termination and depth control. In particular, reducing $D$ may be advantageous for quantum sensing in biological systems, as it could potentially lower microwave-induced heating and improve biocompatibility.

Finally, we calculate the coherence time $T_2^*$ of NV$^-$ center in diamane with F- and O-terminated surface. The decoherence of NV$^-$ center is primarily driven by the interaction with surrounding nuclear spins, typically within a few nanometers. One notable advantage of few-layer diamane is the effective dimensional reduction of the nuclear spin bath to a quasi 2D environment, which can suppress spin dephasing processes. 
In 2-layer F-diamane, the calculated $T_2^*$ = 0.25 $\mu$s. The coherence time decreases with increasing slab thickness, primarily due to the increased number of spin-active $^{13}$C nuclei present in thicker structures. Within a given thickness, $T_2^*$ also exhibits a strong dependence on the depth of the NV center relative to the surface; deeper placement generally leads to longer coherence times, as the coupling to fluctuating surface spins is reduced. 
A Similar trend is observed in O-terminated diamane. Remarkably, in 4-layer O-diamane, the coherence time reaches a maximum of 4.58 $\mu$s when the NV center is located away from the surface. 

\section{Discussion}
\label{sec:discussion}

Diamane thin films inherit the exceptional longitudinal stiffness of bulk diamond and can sustain both in-plane stretching and out-of-plane bending~\cite{gao2018ultrahard, wu2020mechanical}. Recent experiments have further demonstrated that ultrathin diamond membranes possess remarkable flexibility and low surface roughness~\cite{jing2024scalable}.
Such kind of flexibility enables additional tuning methods towards the optical and electronic properties of NV$^-$ center inside and offers a broader scope of applications, such as wearable electronics and diagnostics.
Moreover, the intrinsically high thermal conductivity of diamane~\cite{zhu2019giant} can potentially support long spin-phonon relaxation times for NV$^-$ center.
Despite these advantages, our calculations indicate that ultrathin diamane (2–3 layers) still suffers significantly from surface-related effects. In particular, the surface states can scatter the photoexcited electrons from NV$^-$ center and the scattering rate goes with $\Gamma_{r} = \bracket{\phi_{NV}}{\phi_{s}}$ where $\phi_{NV}$ and $\phi_{s}$ are wavefunctions of NV$^-$ and surface states. This scattering rate is increased in diamane due to the spatial confinement. At the same time, the surface terminations induce large geometry distortion of NV center during excitation, and the corresponding HR factor is larger than it in bulk diamond. In addition, termination species with non-zero nuclear spin—such as H, F, and N—form a magnetic noise environment that contributes to spin decoherence of near-surface NV center. 

The aforementioned drawbacks can be partially mitigated by increasing both the thickness of the diamane film and the depth of the NV center beneath the surface. Our calculations reveal that surface effects are short-ranged and decay rapidly with distance: both the ZPL and HR factor approach bulk-like values when the NV center is located approximately 4~\AA\ below the surface. Given that fluorine and nitrogen possess nearly 100\% natural abundance of non-zero nuclear spin isotopes, isotope engineering to eliminate magnetic noise is practically infeasible. In contrast, oxygen-terminated surfaces are nearly free of nuclear spins, making them ideal for NV-based quantum sensing.
Previous study indicated that the partially oxidized diamond surface with residual H and OH still preserves a clean bandgap without introducing detrimental surface states~\cite{li2019oxygenated,kaviani2014proper}. Another advantage of diamane is that the unwanted impurities created during oxidation are relatively shallow, which makes them easier to be eliminated. The oxygenated surface with etching has been proven to be appropriate to improve $T_2$ because the reduced surface electron spins noise~\cite{favaro2015effect}. We conclude that through elegant control of surface passivation and location of NV center in diamane thin film, the optical and magneto properties of NV center can be optimized compared to its counterpart in diamond. 

\section{Summary and conclusion}
\label{sec:summary}

In summary, we theoretically studied the electronic and optical properties of NV center in few-layer diamane with various surface terminations. Our results show that surface passivation plays a critical role in shaping the band structure of diamane. Specifically, H- and OH-diamane have NEA with image states below CBM, which can mix with the empty states of NV center and cause blinking or bleaching. In contrast, F- and O-diamane provide PEA with clean bandgap without surface states intrusion. 
The size effect of diamane also influence the states of NV center and the corresponding optical excitation.
We find that placing the NV$^-$ center too close to the surface is detrimental to the spin-conserving excitation and leads to reduced charge-state stability. In this regime, the NV$^-$ exhibits an increased HR factor and is more susceptible to ionization into the neutral charge state. However, this surface-induced perturbation is short-ranged and can be effectively suppressed when the NV center is implanted more than 8~\AA\ beneath the surface.
An advantage of diamane lies in its quasi-2D nuclear spin environment, which reduces magnetic noise and extends the coherence time of shallow NV centers. The longest coherence time ($T_2^*$) could be reach to 5 $\mu$s in 2-layer O-diamane, although it gradually decreases with increasing thickness due to the increased $^{13}$C spin bath. Therefore, the depth of the NV center must be carefully optimized to balance optical performance and spin coherence. 
Although a tradeoff is unavoidable, the overall performance of NV$^-$ center in diamane is significantly enhanced compared to its bulk counterpart. Together with the flexibility of diamane, our findings pave the way towards an improved NV-based quantum sensor.  

\section{Methods} 
\label{sec:methods}

In this study, density functional theory (DFT) calculations were performed using the \textit{Vienna ab initio simulation package} (VASP) code~\cite{kresse1996efficiency, kresse1996efficient}, which employs a plane wave basis set. A plane wave cutoff energy of 400~eV was applied. The valence electrons and the core region were described by means of projector augmented wave (PAW) potentials~\cite{blochl1994projector, kresse1999ultrasoft}. The thickness-dependent electronic structure of diamane was investigated for various numbers of carbon bilayers, ranging from 2 to 9, along the (111) diamond surface. The top and bottom surfaces are saturated with same type of terminator to avoid the artificial polarization across the slab. A 12 \AA\ vacuum layer was added in the (111) direction to prevent interactions between periodic slabs. The band structures of surface-passivated diamane were calculated using the Heyd, Scuseria, and Ernzerhof (HSE) hybrid density functional~\cite{heyd2003hybrid}, based on geometries optimized by the generalized gradient approximation (GGA) described by Perdew, Burke, and Ernzerhof (PBE). The hybrid density functional of Heyd, Scuseria, and Ernzerhof (HSE)~\cite{heyd2003hybrid} is used to optimize the geometry and calculate electronic structures of the NV center in diamane. A $4\times4$ supercell model in lateral directions is used to minimize the periodic NV-NV interaction and the $\Gamma$-point sampling scheme is used for the supercell calculation. The convergence threshold for the forces was set to 0.01~eV/\AA. $\Delta$SCF method~\cite{gali2009theory} was employed to calculate electronic excited states. This provide accurate ZPL energy compared with experiment. Due to the size effect, we applied 0.4 eV energy correction on the calculated excitation energies. The zero-field splitting resulting from the dipolar electron - spin electron - spin interaction was calculated within the PAW formalism~\cite{bodrog2013spin} as implemented in VASP by Martijn Marsman. Here the ZFS and hyperfine matrix are evaluated through PBE functional with $6\times6$ supercell model. The hyperfine matrix is calculated with Fermi contact included~\cite{gali2008ab}. 

The electron affinity $\chi$ is the energy difference between the CBM and vacuum level which can be directly extracted from electrostatic potential perpendicular to the surface as 
\begin{equation}
\chi = E_{vac} - E_{CBM} \text{.}
\end{equation}
The NEA induces diamond surface band bending which shifts down the Fermi level and neutralize the NV$^-$. So PEA is required to stabilize the negative charge state of NV.

The formation energy $E_\text{f}$ of NV center is calculated by
\begin{equation}
\begin{split}
E^q_\text{f} = &E^q_\text{NV} - E_\text{slab} -\mu_N +2\mu_C + q\left(E_\text{VBM} + E_\text{Fermi}\right)\\
&+ E_\text{corr}\left(q\right)\text{,}
\end{split}
\end{equation}
where $E_\text{NV}^q$ is the total energy of defective diamane slab with NV at $q$ charge state and $E_\text{slab}$ is the total energy of pure diamane slab. $\mu_N$ and $\mu_C$ are the chemical potential of nitrogen and carbon atom from elementary substances. The Fermi level $E_\text{Fermi}$ represents the chemical potential of electron reservoir and it is aligned to the valence band maximum (VBM) energy, $E_\text{VBM}$. The $E_\text{corr}\left(q\right)$ is the correction term for the charged slab which was done by SXDEFECTALIGN2D code~\cite{freysoldt2020generalized}.

To simulate the decoherence dynamics of near-surface NV$^-$ centers in two-dimensional diamond, we employed the cluster correlation expansion (CCE) method~\cite{zhao2012decoherence, yang2014electron}, a well-established framework for modeling central spin dephasing arising from interactions with a surrounding nuclear spin bath. The coherence time $T_2^*$ under free induction decay (FID) conditions is determined primarily by the static component of the hyperfine field distribution created by nearby nuclear spins.

For nuclei located farther away from the NV$^-$ center, we modeled the hyperfine interaction using the magnetic dipole-dipole approximation. In this region, extending up to a $120 \times 120$ supercell, the hyperfine tensors were calculated analytically based on nuclear positions. This hybrid approach balances computational efficiency with accuracy, enabling the simulation of large spin baths while preserving detail near the defect. 

The nuclear spin bath predominantly consists of $^{13}$C nuclei ($I = 1/2$, natural abundance 1.1\%) randomly distributed on lattice sites. Surface nuclear spins with nonzero spin—specifically $^{19}$F ($I = 1/2$, 100\% abundance) and $^{17}$O ($I = 5/2$, 0.037\% abundance)—were included when relevant to capture decoherence effects from surface termination. The spin bath size within the $120 \times 120$ supercell was sufficiently large to ensure convergence of the coherence function. To account for statistical variations, 1000 random nuclear spin configurations consistent with the isotopic abundances were generated for ensemble averaging.

The CCE method was then applied to compute the electron spin coherence function $L(t)$ under FID conditions. The coherence decay was obtained by averaging over the ensemble of nuclear spin configurations. The envelope of $L(t)$ follows an approximately Gaussian decay, $L(t) \sim \exp[-(t/T_2^*)^2]$, from which the dephasing time $T_2^*$ was extracted.

\section*{Competing interests}
The authors declare that there are no competing interests.

\section*{Data Availability}
The data that support the findings of this study are available from the corresponding author upon reasonable request.

%
%
\begin{acknowledgements}
BH acknowledges the National Key Research and Development of China (Grant No.\ 2022YFA1402400), NSFC (Grant No.\ 12088101), and NSAF (Grant No.\ U2230402). PL acknowledges the NSFC (Grant No.\ 12404094). This research was supported by Open Research Fund of CNMGE Platform $\&$ NSCC-TJ.

\end{acknowledgements}

\bibliography{name.bib}

\end{document}